\documentclass[12pt]{article}
\usepackage{epsfig}
\def\gamgam{\mbox{$\gamma \gamma$}}  
\newcommand{\ra}{\rightarrow}

\def\be{\begin{equation}}
\def\ee{\end{equation}}

\begin{document}

\begin{flushright}
IISc-CTS/8/03\\
hep-ph/0311185\\
\end{flushright}
\begin{center}

{\Large
{\bf    CP property of the Higgs at the $\gamgam$ colliders using  
$t \bar t$ production.}\footnote{Talk presented at the 8th Accelerator 
and Particle Physics Institute, APPI2003, Feb.25-28,2003, Appi, Japan.} \\[5ex]
R.M. Godbole\footnote{e-mail:rohini@cts.iisc.ernet.in}

{\it Centre for Theoretical Studies, Indian Institute of Science, Bangalore, 560 012, India.}}
\end{center}
\vspace{1.5cm}
{\begin{center} ABSTRACT \end{center}}
\vspace{-4truemm}
{\small{
We present results of an investigation to study CP violation in the
Higgs sector in  $t\bar t$ production at a $\gamma\gamma$-collider, via the 
process $ \gamgam \ra \phi \ra t \bar t$ where the $\phi$ is a scalar with
indeterminate CP parity. The study is performed in a model independent way 
parametrising the CP violating couplings in terms of six form factors
$\{\Re(S_{\gamma}), \Im(S_{\gamma}), \Re(P_{\gamma}),
\Im(P_{\gamma}), S_t, P_t\}$.  The CP violation is reflected 
in the polarisation asymmetry of the produced top quark. We use the 
angular distribution of the decay lepton from $t/\bar t$ as a diagnostic of 
this polarisation asymmetry and hence of the CP mixing, after showing  that 
the asymmetries in the angular distribution are indpendent of any CP violation 
in the $tbW$ vertex.  We construct combined asymmetries
in the initial state lepton (photon) polarization and the final state lepton
charge and study how well different combinations of these form factors
can be probed by measurements of these asymmetries, using circularly polarized
photons. We demonstrate the feasibility of the  method to probe CP 
violation in the Higgs sector at the level induced by loop effects in 
supersymmetric theories, using  realistic photon spectra expected for a TESLA
like $e^+ e^-$ collider. We investigate  the sensitivity of our method for 
for different widths of the scalar as well as for the more realistic 
backscattered laser photon spectrum resulting from the inlcusion of the 
nonlinear effects.  }}

\newpage  

\section{Introduction}
\label{intro} 
While the standard model (SM) has been proved to provide the correct
description of all the fundamental particles and their interactions, direct
experimental verification of the Higgs sector and a basic understanding
of the mechanism for the generation of the observed CP violation is
still lacking.  Many models with an extended Higgs sector have  CP violation
in the Higgs sector.
In this context  there are then two important questions that need to be
answered viz., if CP is conserved in the Higgs sector, how well can the
CP transformation properties of the, possibly more than one, neutral Higgses
be established. If it is violated then one wishes to study how is this 
CP violation reflected in Higgs mixing as well as couplings and how well can 
these be measured at the colliders. 
CP violation in the Higgs sector can be either explicit, spontaneous or 
loop-induced. The last has been studied in great detail in the context 
of the minimal Supersymmetric Standard Model (MSSM) recently~\cite{mssmcp} 
and arises from loops containing sparticles and nonzero phases of the MSSM 
parameters $\mu$ and $A_t$.

$\gamgam$ colliders will make possible an accurate measurement of the width 
of a Higgs scalar into  $\gamma \gamma$~\cite{stephan} and $WW/ZZ$~\cite{maria} 
channels. A study of the former can give very important information about the 
physics beyond standard model (SM) due to the nondecoupling nature of this 
width. Further, $\gamgam$ collisions will also offer the possibility of a study of heavy neutral Higgses $H/A$ of the MSSM through their production in 
$\gamgam$ collisions, followed by their decay into a pair of neutralinos, 
thus making possible an exploration of the MSSM Higgs sector in a region 
of the parameter space not accessible to the LHC~\cite{zerwas}.  

Photon Colliders with their democratic coupling to both the 
CP even and the CP odd scalars and the possibility of polarised photon beams, 
offer the best chance to explore the CP property of the scalar sector. Using
$\gamgam $ colliders with linearly polarised photon beams, it is possible
to study the CP property of the Higgs from just the polarisation dependence of
the cross-section~\cite{gunion}.  The $ZZ$ decay can also be used very 
effectively to make a model independent determination of the CP nature of the
Higgs boson at the $e^+e^-$, \gamgam\ colliders  as well as the 
LHC/cite{david}. It has 
been shown~\cite{asakawa} that even in the case of photon colliders with just 
the circular polarisation, it might be possible to probe the CP property of 
the Higgs by looking at the net polarisation of the top quarks produced in the 
process  $\gamma \gamma \rightarrow \phi \rightarrow t \bar t$.  With linearly 
polarised  $\gamma$ and the $t \bar t$ decay of the scalar it should be 
possible to completely reconstruct the $\phi \gamma \gamma$ and the 
$\phi t \bar t$ vertex, using the resulting polarisation  asymmetries of the 
$t$. The $t$ quark being very heavy decays before it hadronises. Hence the
decay lepton energy and angular distributions can be used as an analyser of 
the $t$ polarisation~\cite{saupol}.  Thus a study of the simple inclusive 
lepton angular distributions in the 
$\gamma \gamma \rightarrow \phi \rightarrow t \bar t$.
can yield information about the CP property of the $\phi$. We considered
$\gamma \gamma$  production of a  $t \bar t$ pair through the s-channel 
exchange of a scalar $\phi$ of indeterminate CP property and studied~\cite{us}
how well the CP property of such a scalar can be probed  using  mixed 
asymmetries with respect to the final state lepton charge and initial state 
photon polarisation. Our analysis considered only the case of circularly 
polarised lasers. Recently, a calculation~\cite{kaorunew}
of  helicity amplitudes for the 
consequent lepton production coming from the $t$ decay has been 
performed including the case of the linear polarisation of the laser photon.

In sections 2 and 3, I recapitulate the notation, the 
methodology, along with  a discussion of the independence of the decay lepton 
angular distribution from any anomalous $tbW$ vertex. I end with an example of 
the sensitivity expected for a particular point in the MSSM parameter space 
at a $\gamgam $ collider with ideal backscattered laser photon 
spectrum\cite{ginzburg}.  In the last section,   I then present update of 
these results  using the parametrisation~\cite{zarnecki} of a more realistic 
backscattered laser spectrum resulting from inclusion of nonlinear 
effects~\cite{telnov} and that of a variation in the width of the scalar.

\section{Formalism and calculation of the decay $l$ angular distribution}
\begin{figure}[htb]
\begin{center}
\includegraphics*[scale=1.0]{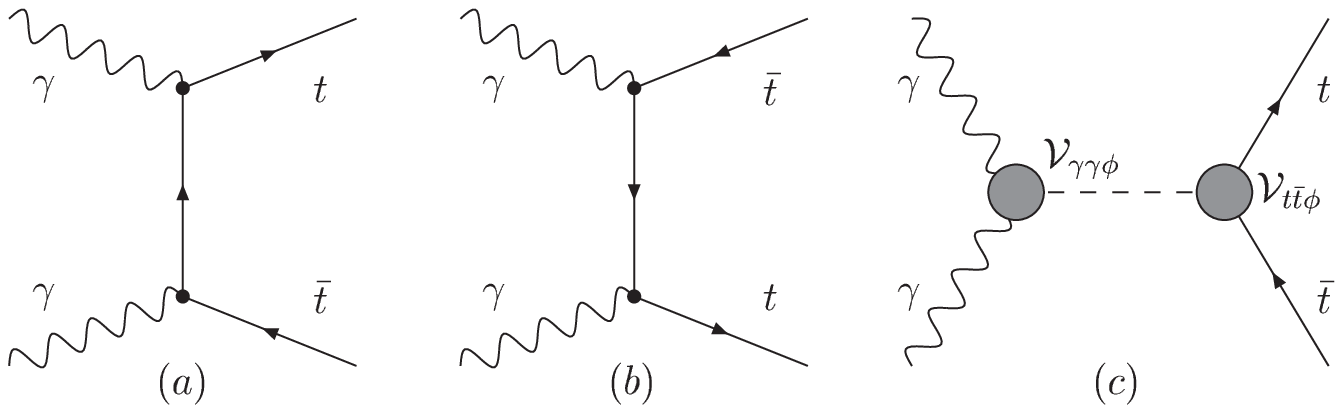}
\caption{Feynman diagrams for the $t \bar t$ production in $\gamma \gamma$ 
collisions.}
\label{fig1}
\end{center}
\end{figure}

The process we study is shown in Fig.~\ref{fig1}.
The diagrams shown in Figs.~\ref{fig1} (a) and  \ref{fig1} (b) give what we 
will deonote later as the SM contirbution and   Fig.~\ref{fig1} (c) shows
the contribution from the  production of a  scalar $\phi$ with indeterminate
CP parity. The most general model independent expression for the 
$\phi \gamma \gamma$ and $\phi t \bar t$ vertices for such a $\phi$ that we use
is written as, 
\begin{eqnarray}
&{\cal V}_{t\bar t \phi}  =  -ie\frac{m_t}{M_W} \left(S_t+i\gamma^5P_t\right),\\
&{\cal V}_{\gamma\gamma\phi}  = \frac{-i\sqrt{s}\alpha}{4\pi}\left[S_{\gamma}(s)
\left(\epsilon_1.\epsilon_2- {2\over s} (\epsilon_1.k_2)(\epsilon_2.k_1) \right)
\right.
-\left. {2 P_{\gamma}(s) \over s}  \epsilon_{\mu\nu\alpha\beta}\epsilon_1^{\mu}
\epsilon_2^{\nu} k_1^{\alpha} k_2^{\beta} \right]. 
\end{eqnarray}
Here $k_1$ and $k_2$ are the four-momenta of colliding photons and 
$\epsilon_{1,2}$ are the photon polarisation vectors. Form factors 
$S_t, P_t$ can be taken to be  real without any loss of generality 
whereas  the most general $S_\gamma, P_\gamma$ are required to be complex. 
Simultaneous presence of nonzero $P,S$ coupling implies CP violation. Of course
in a given model the predictions for $P_t,S_t$ and $P_\gamma, S_\gamma$
are correlated. The most general $tbW$ vertex  can be written as,
\begin{eqnarray}
\Gamma^{\mu}_{tbW}&=-\frac{g}{\sqrt{2}}V_{tb}\left[\gamma^{\mu}
\left(f_{1L}P_L + f_{1R} P_R\right)\right.
-\left.\frac{i}{M_W}\sigma^{\mu\nu}(p_t-p_b)_{\nu}
\left(f_{2L}P_L + f_{2R} P_R\right)\right]\\
{\bar\Gamma}^{\mu}_{tbW}&=-\frac{g}{\sqrt{2}}V_{tb}^*\left[\gamma^{\mu}
\left({\bar f}_{1L}P_L + {\bar f}_{1R} P_R\right)\right.
-\left.\frac{i}{M_W}\sigma^{\mu\nu}
(p_{\bar t}-p_{\bar b})_{\nu}
\left({\bar f}_{2L}P_L + {\bar f}_{2R} P_R\right)\right]
\label{tbw}
\end{eqnarray}
Following  LEP measurements we take  $f_{1L} ={\bar f}_{1L} =1 $. All the other
$f_i, \bar f_i$ are necessarily small. $f_{2R}, {\bar f}_{2L}$ are the only 
nonstandard part of the $tbW$ vertex that contribute to the  angular and
energy distribution of the lepton in the limit $m_b =0$. In our analysis we 
keep only terms linear in them.  We calculate the analytical expression for 
helicity amplitudes and the
differential cross-section for $\gamma \gamma \rightarrow t \bar t
\rightarrow l^+ b \nu_l \bar t$ and  the angular distribution of the consequent 
decay lepton using the general $\phi t \bar t, \phi \gamma \gamma$ and 
the $tbW$ vertices given above.

\noindent{\bf Independence of the $l$ angular distribution from the anomalous
part of the $tbW$ vertex.}

The expression for the differential distribution for the decay lepton $l$ can
be written as
\begin{equation}
\frac{d\sigma}{d\cos\theta_t \ d\cos\theta_{l^+} \ dE_{l^+} \ d\phi_{l^+}}=
\frac{3e^4 g^4 \beta E_{l^+}}{64(4\pi)^4 s \Gamma_tm_t \Gamma_WM_W}
\sum_{\lambda,\lambda'} \underbrace{\rho'^+(\lambda,\lambda')}_{c.m. \ frame}
\underbrace{\left[\frac{\Gamma'(\lambda,\lambda')}{m_t E_{l^+}^0}\right]}_{rest
\ frame}.
\label{diffdistr}
\end{equation}
In the above, $E_{l^+}^0$ is energy of $l^+$ in the rest frame of $t$ quark.
The production and decay density matrices $\rho^+(\lambda,\lambda'),
\Gamma(\lambda,\lambda')$ are given by
\begin{eqnarray}
\rho^+(\lambda,\lambda') = e^4\rho'^+(\lambda,\lambda') = \sum
\rho_1(\lambda_1,\lambda_1')\rho_2(\lambda_2,\lambda_2') 
{\cal M}(\lambda_1,\lambda_2,\lambda,\lambda_{\bar t})
{\cal M}^*(\lambda_1',\lambda_2', \lambda',\lambda_{\bar t})\\
\Gamma(\lambda,\lambda') = g^4 |\Delta(p_W^2)|^2  \ \Gamma'(\lambda,
\lambda') = \frac{1}{2\pi}\int d\alpha 
\times \sum
M_{\Gamma}(\lambda,\lambda_b,\lambda_{l^+},\lambda_{\nu}) \
M_{\Gamma}^*(\lambda',\lambda_b,\lambda_{l^+},\lambda_{\nu})
\label{helamp}
\end{eqnarray}

In the above, $\alpha$ is the  azimuthal angle of $b$-quark in the rest-frame 
of $t$-quark with $z$-axis pointing in the direction of momentum of lepton and 
$\rho_{1(2)}$ are the photon density matrices.  The decay density matrix 
elements are in the $t,\bar t$ frame. For example the $+,+ (-,-)$ element
is given by,
\begin{equation}
\Gamma(\pm,\pm) = g^4m_tE_{l^+}^0|\Delta_W(p_W^2)|^2 \ (m_t^2-2p_t.p_{l^+
}) (1\pm\cos\theta_{l^+})\left(1+\frac{\Re(f_{2R})}{\sqrt{r}}
\frac{M_W^2}{p_t.p_{l^+}}\right)
\end{equation}

The decay $l$ angular distribution can be obtained analytically by integrating 
Eq.~\ref{diffdistr} over $E_l,\cos\theta_t$ and $\phi_l$. We find that  the 
only effect of the anomalous part of the $tbW$ coupling on $l$ angular 
distribution is an overall factor $1+2r-6\Re(f^\pm)\sqrt{r}$ {\it independent} 
of any kinematical variables. We further find that the 
total width of $t$-quark calculated upto linear order in the anomalous
vertex factors receives the {\it same} factor. As a result
the angular distribution of the decay lepton  is unaltered by the 
anomalous part of the $tbW$ couplings to the linear approximation. Since the
correlation between the top spin and the angle of emission of the decay lepton
is essentially a result of the  $V-A$ nature of the $tbW$ coupling, it can
be thus used as a true polariometer for the polarisation of the $t$. The 
energy distribution of the decay lepton, which also reflects the polarisation 
of the parent $t$ quark does get affected by the presence of the anomalous part
of the $tbW$ couplings.  Thus the angular  distribution of the decay $l$ is  
a very interesting observable for which the only source of the CP
violating asymmetry will then be the production process. Further, the 
construction of CP violating asymmetries using the angular distributions of the
decay $l$ does not need precise reconstruction of the top rest frame and the 
consequent prescise knolwedge of the  top quark momentum. 
For the case of $e^+e^- \rightarrow t \bar t$ followed by subsequent
$t/\bar t$ decay,  this was observed  earlier \cite{hioki,sdr}.  It was proved
recently by two groups independently; for a two-photon initial state by
Ohkuma~\cite{ohk}, for an arbitrary two-body initial state in \cite{grza}
and further keeping $m_b$ non-zero in \cite{grza1}.  These latter 
derivations use the method developed by Tsai and collaborators \cite{tsai}
for incorporating the production and decay of a massive spin-half particle.
Our current derivation made use of helicity amplitudes and provides an 
independent verification of these results.

\section{Asymmetries and their sensitivity to CP violation expected 
due to loop effects}

The cross-section has a nontrivial dependence on the poalrisation of the 
initial state photons as $\phi$ exchange diagram contributes only when both
colliding photons have same helicity due to its spin $0$ nature. Further,
SM contribution is peaked in the forward and backward direction whereas the
scalar 
exchange contribution is independent of the production angle $\theta_t$.
Hence angular cuts to redcue the SM contribution along with choice of equal 
helicities for both the colliding photons can maximise polarisation asymmetries for the produced $t \bar t$ pair, giving a better measure of the $CP$ violating
nature of the $s-$ channel contribution. Another thing to note is that the 
polarisation of the collidiing photon is decided by the polarisation of the 
initial lepton and that of the  laser photon used in the backscattering which
gives rise to the energetic colliding photon. 
\begin{figure}[htb]
\begin{center}
\includegraphics*[scale=0.8]{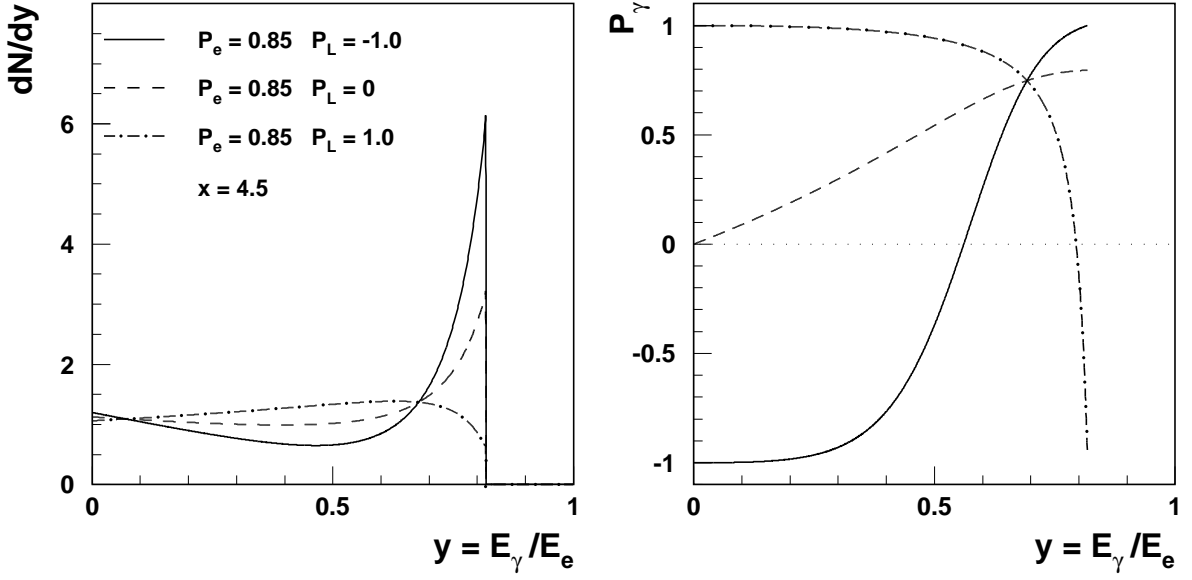}
\caption{Energy spectrum and the polarisation expected for the backscattered 
laser for ideal case ~\protect\cite{ginzburg}} 
\label{fig2}
\end{center}
\end{figure}
Fig.~\ref{fig2} shows the energy spectrum and the polarisation expected for 
the backscattered laser for the ideal case ~\protect\cite{ginzburg}, i.e., 
neglecting the nonlinear effects, for different choices of the $e$ and the 
laser polarisation. One can choose  $\lambda_e\lambda_l=-1$ 
to get a hard photon spectrum. Further, as discussed above,  one sets 
$\lambda_{e^-} = \lambda_{e^+}$ to maximise the sensitivity to
possible $CP$ violating interactions coming from the scalar exchange. Thus
all the polarisations are fixed wrt to that of one of the 
leptons.\footnote{For sake of definiteness we use the case of a parent $e^+e^-$
collider. But all the discussion applies equally well to the case of an 
$e^- e^-$ collider.} Thus there are two choices of the 
initial state lepton polarisation, $\lambda_{e^-}= \lambda_{e^+} = +1 $
and $-1$.  In the final state one can look for either $l^+$ or $l^-$.
This makes four possible combination of  cross-sections depending on the
initial state photon and the final state $l$ charge: $\sigma(+,+), 
\sigma(+,-), \sigma(-,+)$ and $ \sigma(-,-)$. CP conservation will imply,
for example for the QED contribution which we call the SM contribution, 
$\sigma(+,+) = \sigma(-,-)$. Using these now we  construct asymmetries which 
will be sensitive to $\phi$ coupling.   

\noindent{\bf Asymmetries}

For defining asymmetries, we choose two polarised cross-section at a time
out of four available, and can define six asymmetries as,

\begin{eqnarray}
{\cal A}_1  =  \frac{\sigma(+,+)-\sigma(-,-)}{\sigma(+,+)+\sigma(-,-)};
{\cal A}_2  =  \frac{\sigma(+,-)-\sigma(-,+)}{\sigma(+,-)+\sigma(-,+)};
{\cal A}_3  =  \frac{\sigma(+,+)-\sigma(-,+)}{\sigma(+,+)+\sigma(-,+)}\nonumber\\
{\cal A}_4  =  \frac{\sigma(+,-)-\sigma(-,-)}{\sigma(+,-)+\sigma(-,-)};
{\cal A}_5  =  \frac{\sigma(+,+)-\sigma(+,-)}{\sigma(+,+)+\sigma(+,-)};
{\cal A}_6  =  \frac{\sigma(-,+)-\sigma(-,-)}{\sigma(-,+)+\sigma(-,-)}
\label{asymm}
\end{eqnarray}

${\cal A}_5$ and ${\cal A}_6$ are charge asymmetries  for a given polarisation.
These will be zero if $\theta_0 \rightarrow 0$. The same is not true of course
of the purely CP violating ${\cal A}_1$ and ${\cal A}_2$. ${\cal A}_3$ and 
${\cal A}_4$ are the polarisation asymmetries for a given lepton charge. The
phenomenon of nonvanishing charge asymmetries even for the SM case, for 
polarised photon beams has been also been observed recently in the context 
of $\mu+ \mu-$ pair production\cite{ginzburgnew}. However these can not be
directly compared as we have constructed the asymmetries in terms of the 
polarisation of the incoming lepton beam rather than that of the photon.
The contribution of the $s$ channel diagram to the asymmetries can be 
enhanced by the choice of relative polarisation of the $e^+ e^-$ beams and the 
angular cuts as mentioned above as well as that of the beam energy. Of 
course only three of the  asymmetries given above are linearly independent 
of each other. The sensitivity of the these  asymmetries to the various 
couplings of the scalar $\phi$ in general and to the CP violating part in 
particular, can be best judged by taking a specific numerical example.

To that end we choose the values of the form factors obtained in the second of
Ref. \cite{asakawa} for $\tan \beta =3$, with all sparticles heavy and maximal 
phase: $m_\phi  = 500 GeV , \Gamma_\phi = 1.9 GeV, S_t =  0.33 , P_t =  0.15,
S_{\gamma}  =  -1.3-1.2i , P_{\gamma}  =  -0.51+1.1i.$ 
\begin{figure}[htb]
\begin{center}
\includegraphics*[scale=0.8]{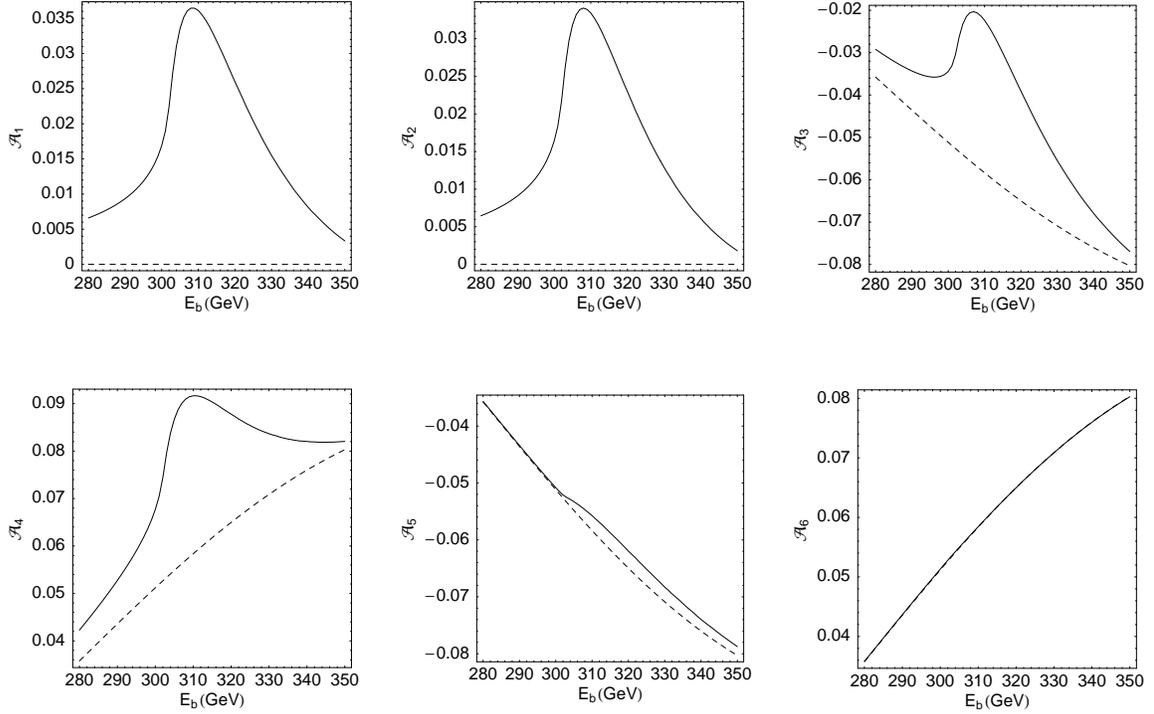}
\caption{Asymmetries expected for the SM cotribution (continuum) and
for the  chosen MSSM point (solid line)}
\label{fig3}
\end{center}
\end{figure}
We notice that the expected asymmetries are not  insubstantial. Even the 
CP violating asymmetries are at the level of $3$--$4$ \%.
The results presented correspond to a cut on the lepton angle 
$\theta_0 = 60^\circ$.  The choice of the cutoff angle on the lepton and 
also of the energy was optimised by studying the sensitivity of a particular 
asymmetry measurement.
The number of events corresponding to the asymmetry are ${\cal L }
\Delta \sigma$.  For the asymmetry to be measurable say at $95\%$ CL,
we must have at least ${\cal L} \Delta \sigma > 1.96 \sqrt{{\cal L} \sigma}$. 
A measure of the senstivity thus can be
$${{{\cal L} {\Delta \sigma}}\over {1.96 \sqrt{{\cal L}\sigma}}}  
= {\sqrt{{\cal L}}\over 1.96} \times {{\Delta \sigma} \over {\sqrt{\sigma}}}$$
The larger the asymmetry as compared to the fluctuations, the larger the 
sensitivity with which it can be measured.  We define {\em sensitivity} as,
${\cal S}=\frac{{\cal A}}{\delta{\cal A}} \propto \frac{\Delta\sigma}
{\sqrt{\sigma}}$.  
For this choice of the scalar mass and the ideal backscattered photon,
the cross-sections, the asymmetries and the sensitivity is optimised 
by choosing $E_b = 310$ GeV and two choices of angular cuts $20^\circ$
and $60^\circ$.  With this choice of energy and the ideal Ginzburg 
spectrum, we then anlaysed  how well various scalar couplings can be studied
using these asymmetries.
\vspace{1cm}

{\bf Analysis of the sensitivity of these asymmetries to the scalar couplings:}

$CP$ properties of the Higgs determined if we know all the {\it four}
form-factors $S_t,  P_t,$ $ \Re(S_{\gamma}),  \Im(S_{\gamma}),$
and $\Re(P_{\gamma}),  \Im(P_{\gamma}) $. They appear in the production density 
matrix in eight combinations,  $x_i$ and $y_i$, $(i=1,...4)$ given by;
\begin{eqnarray}
x_1 = S_t\Re(S_{\gamma}), \ x_2 = S_t\Im(S_{\gamma}), \ x_3 = P_t\Re(P_{\gamma}), \ x_4 =  P_t\Im(P_{\gamma})\label{x}\\
y_1 = S_t\Re(P_{\gamma}), \ y_2 =S_t\Im(P_{\gamma}), \ y_3 = P_t\Re(S_{\gamma}),\ y_4 =  P_t\Im(S_{\gamma}).
\label{y}
\end{eqnarray}  
$x_i$ given by Eq. ~\ref{x} all being CP even and $y_i$ of Eq. ~\ref{y} 
all CP odd. Only five of these are linearly independent and we have,
\begin{equation}
y_1.y_3  = x_1.x_3,  \ y_2.y_4 = x_2.x_4, \ y_1.x_4  = y_2.x_3,  
y_4.x_1 = y_3.x_2.
\label{constraints}
\end{equation}

Since all the asymmetries are functions of $x_i,y_i, i=1,4$ one would like to
explore the sensitivity of the asymmetry measurements to values of $x_i,y_i$. 
If for certain values of the form-factors the asymmetries lie within the
fluctuation from their SM values, then that particlar point in the
parameter space cannot be distinguished from SM at that luminosity. Using this
as the criterion we can identify regions in the $x_i$--$y_j$ plane where it is
possible to probe a particular non-zero value of $x_i,y_j$ and hence probe
the deviation from the SM amplitude. The region where this is not possible can
be termed as the blind region of the particular asymmetry being considered.
Thus the set of parameters \{$x_i,y_i$\} will be inside the
blind region at a given luminosity if,
$$|{\cal A}(\{x_i,y_i\})-{\cal A}_{SM}| \leq \delta{\cal A}_{SM} =
\frac{f}{\sqrt{\sigma_{SM} L}}\sqrt{1+{\cal A}_{SM}^2}.$$
It is clear that using just the three linearly independent asymmetries it
will not be possible to extract all the  $x_i,y_i$  and hence all the
form factors involved, Instead we take only two of the eight $x_i,y_j$ nonzero 
at a time and ask how well one can constrain these using the measurements of 
the asymmetries at a given luminosity. We study blind regions in the various 
$x_i$--$y_j$ planes  for all the different asymmetries and choose the best one
in each case. The analysis is performed for the choice of the beam energy 
$E_b$ and angular cut  $\theta_0$ mentioned above. 

\begin{figure}[htb]
\rotatebox{270}{ \scalebox{.7}{ \includegraphics{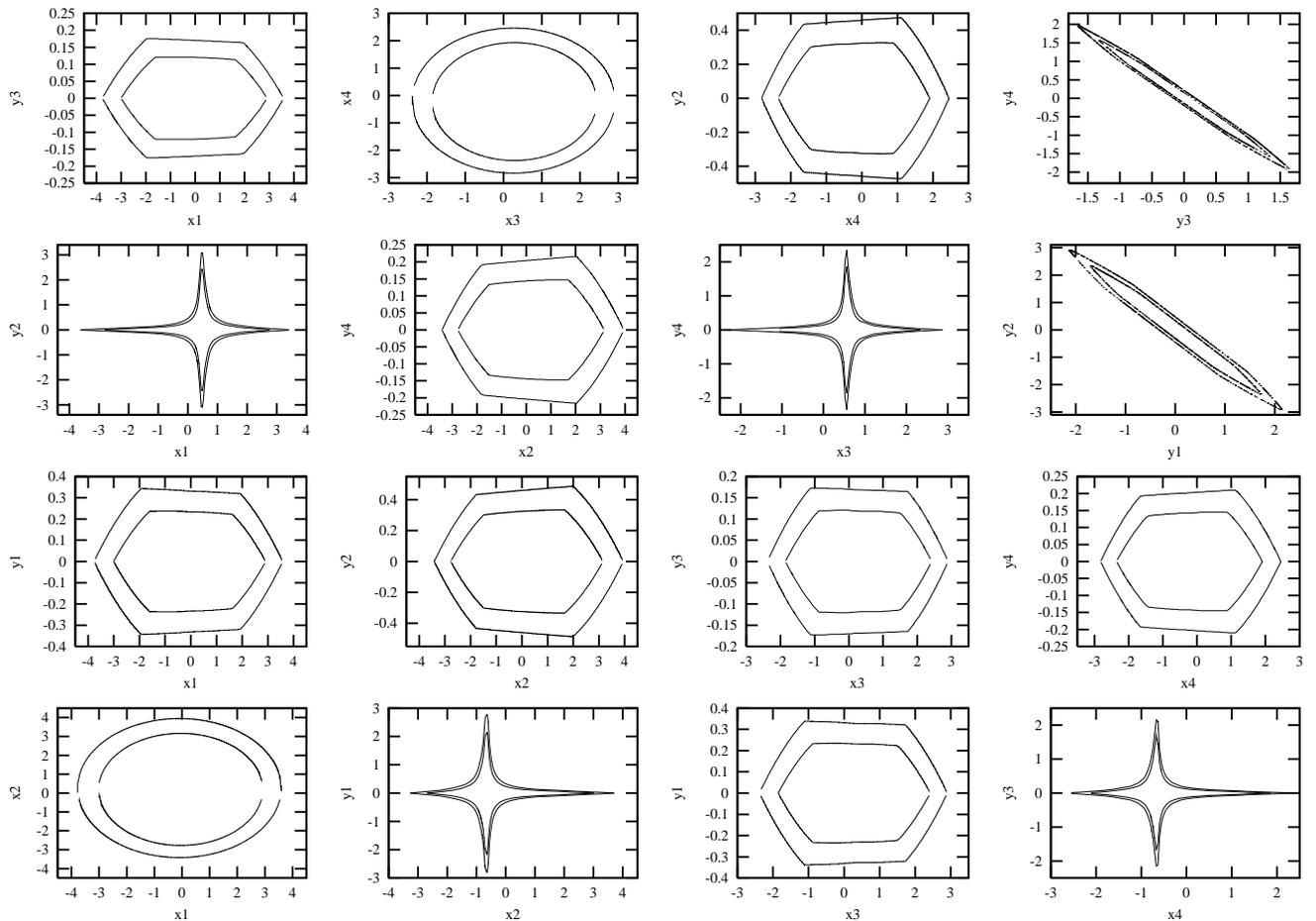} } }
\caption{Blind regions in the various $x_i$--$y_j$ planes, the larger 
and smaller region corresponding to  luminosity $ 500 {\rm fb}^{-1}$,
$1000 {\rm fb}^{-1}$ respectively.  } \label{limits}
\end{figure}
Fig.~\ref{limits} shows the blind regions in the various $x_i$--$y_j$ planes;
the larger and smaller region corresponding to a luminosity of 
$ 500 {\rm fb}^{-1}, $ and  $1000 {\rm fb}^{-1}$ respectively.
Thus we see that these asymmetries do indeed have the potential of probing 
nonzero values of $y_i$ and hence probing the CP violation in the
Higgs sector. 

\begin{table} 
\begin{tabular}{|c|cc|cc|c||}
\hline
&&&&&\\
  &min         &max           &min             &max  &MSSM     \\
  &(500 fb$^{-1}$)&(500 fb$^{-1}$)&(1000 fb$^{-1}$)&(1000 fb$^{-1}$)&value\\
\hline\hline
$x_1$&  $-3.775 $&        3.594 &         $-2.990 $&       2.869 &$-0.429$\\
$x_2$&  $-3.413 $&        3.896 &         $-2.748 $&       3.111 &$-0.396$\\
$x_3$&  $-2.386 $&        2.873 &         $-1.842 $&       2.386 &$-0.077$\\
$x_4$&  $-2.837 $&        2.465 &         $-2.375 $&       1.930 &$+0.165$\\
$y_1$&  $-2.786 $&        2.786 &         $-2.148 $&       2.148 &$-0.168$\\
$y_2$&  $-3.095 $&        3.095 &         $-2.433 $&       2.433 &$+0.363$\\
$y_3$&  $-2.155 $&        2.155 &         $-1.687 $&       1.687 &$-0.195$\\
$y_4$&  $-2.346 $&        2.346 &         $-1.867 $&       1.867 &$-0.180$\\
\hline
\end{tabular}
\caption{Limits possible on $x_i,y_j$ for two different luminosities at 
$95 \% $ CL}
\label{table1}
\end{table}

The results of Fig.~\ref{limits} can also be summarised in  terms of limits
upto which the $x_i,y_j$ can be probed using the  measurements of asymmetries
alone, if all of them are allowed to vary simultaneously. These are given in 
Table~\ref{table1}.
The last column gives  values expected for the chosen MSSM point.
It is clear that the limits that the asymmetries can put if all of the $x_i,y_j$ are allowed to vary simultaneously are not very good. But two things should be
noted here. Firstly one is sugeesting this as a second generation experiment
after the Higgs discovery, hence one might be able to use the known partial
information on $x_i,y_j$ and thus be able to constrain these combinations and 
hence the form factors better. Secondly, it has been shown that with the
use of linearly polarised photons one can indeed  reconstruct all the form 
factors completely~\cite{asakawa,kaorunew}. It would be interesting to extend 
our analysis to the case of the linearly polarised photons as well.

{\bf Discrimination between the SM and the MSSM}
 
We also investigated  the confidence level with which the particular  MSSM 
point chosen by us can be discriminiated from the  QED SM background using the 
asymmetries alone. To that end we calculated  the $x_i,y_j$ for our  choice of
CP violating parameters given by the MSSM point and then investigated the 
geometry of the blind regions about this point. Again we varied a pair of 
$x_i,y_j$ at  a time, keeping all the other fixed at the values expected for 
the MSSM point.  We thus obtained the blind regions around the point the same 
way as we did for the SM. 
\begin{figure}
\begin{center}
\includegraphics*[scale=0.7]{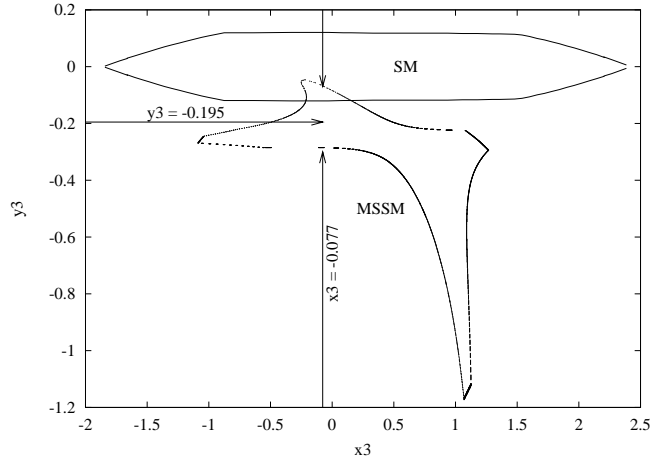}
\label{smmssm}
\caption{Blind regions for the chosen MSSM point and that for the plain SM QED
contribution for $1000 {\rm fb}^{-1}$ at $95 \%$ CL.}
\end{center}
\end{figure}
Fig. ~\ref{smmssm} shows that the blind regions for the SM point and around the
MSSM point have very little overlap. This demonstrates that the method is
indeed sensitive to the CP violation at the level produced by loop effects.

\section{Effect of higher Higgs width and more realistic photon spectra}
It is obvious that the sensitivity will depend critically on the width of the 
scalar and is likely to decrease as the width increases. Further, the energy
spectrum of the backscattered laser photon and more importantly the 
polarisation is subject to large effects from multiple interactions.
\begin{figure}
\begin{center}
\includegraphics*[scale=0.6]{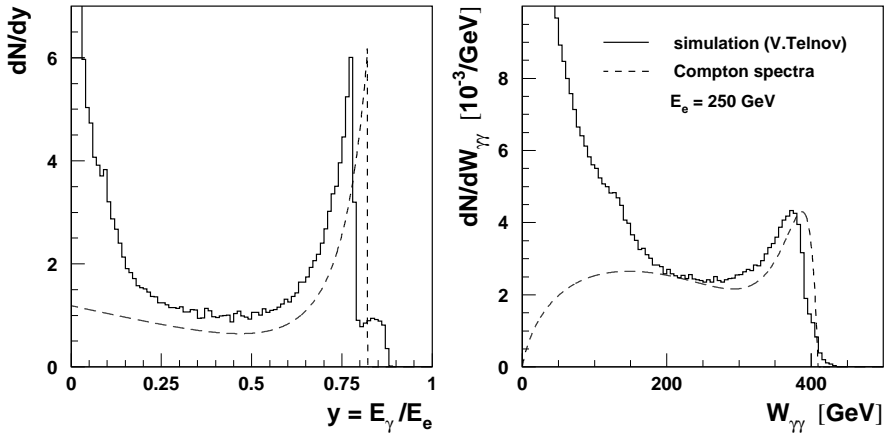}
\label{nonlinear}
\caption{The energy spectrum for the single photon and the spectrum in the two 
photon invariant mass $W_{\gamma \gamma}$ for the backscattered laser 
photons after inclusion of the nonlinear effects as parametrised in 
\protect\cite{zarnecki} and compared with simulation~\protect\cite{telnov}, for
$E_{beam} = 250 $ GeV.}
\label{zarspec}
\end{center}
\end{figure}
\begin{figure}
\begin{center}
\includegraphics*[scale=0.6]{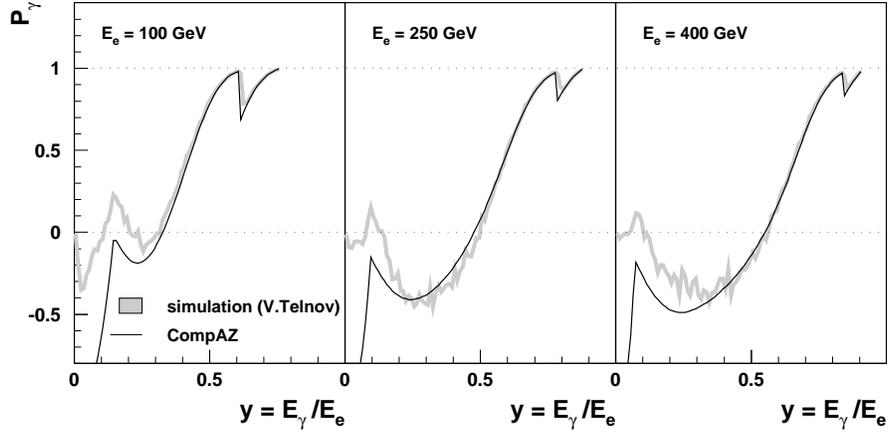}
\caption{The polarisation $P_\gamma$ for the backscattered laser photon 
as a function of the fraction of the beam energy carried by the photon,
after inclusion of the nonlinear effects as parametrised in
\protect\cite{zarnecki} and compared with simulation~\protect\cite{telnov}, for
$E_{beam} = 100,250$ and $400$ GeV.}
\label{zarpol}
\end{center}
\end{figure}
The latter is illustrated in Figs.~\ref{zarspec}, \ref{zarpol} taken from
Ref.~\cite{zarnecki}. Fig.~\ref{zarspec} shows the The energy spectrum for the 
single photon and the spectrum in the two
photon invariant mass $W_{\gamma \gamma}$ for the backscattered laser
photons after inclusion of the nonlinear effects, as parametrised in
\protect\cite{zarnecki} and compared with simulation~\protect\cite{telnov}, 
for $E_{beam} = 250 $ GeV. Fig.~\ref{zarpol} shows the expected polarisation
for three different values of the beam energy. This is to be compared with 
right panel of Fig.~\ref{fig2}.
Since the asymmetries depend crucially on the polarisation it is likely that
our study of sensitivity will get affected by this change in the spectrum
and the polarisation. Further, for this more realistic case the $e^-$ is taken
to have only $85\%$ polarisation as opposed to the $100 \%$ assumed in our 
earlier study. Fig.\ref{asmcompaz} shows four of the
asymmetries of Eq.~\ref{asymm} obtained using the  
CompAZ parametrisation~\cite{zarnecki} of the more realistic 
spectra~\cite{telnov}, plotted as a function of $E_b$ .
\begin{figure}
\begin{center}
\includegraphics*[scale=0.6]{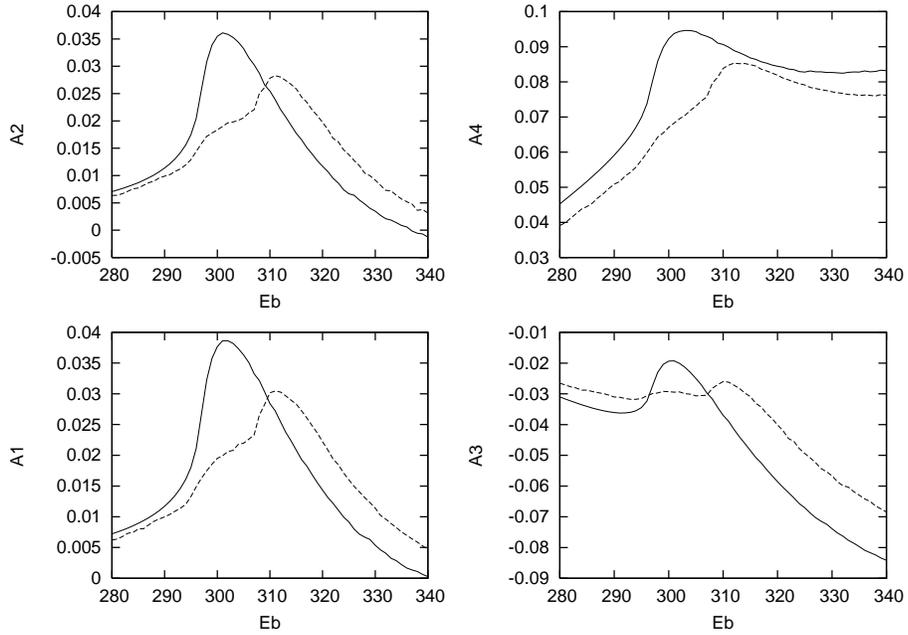} 
\caption{Four of the asymmetries of Eq.~\protect\ref{asymm} for the 
ideal and for the realistic spectrum.}
\label{asmcompaz}
\end{center}
\end{figure}
For the more realistic spectrum there is also a net decrease in the effective
luminosity as the multiple interactions increase the number of the photons 
in the low energy region. The major effect of the use of the more realistic
spectrum seems to be this decrease in the luminosity of the useful, energetic
photons.

Even though we have performed our studies in a model independent way, we are 
specifically also interested in the case of a MSSM scalar. The heavy  MSSM 
scalar is not expected to be very wide. 
\begin{figure}
\begin{center}
\includegraphics*[scale=0.4]{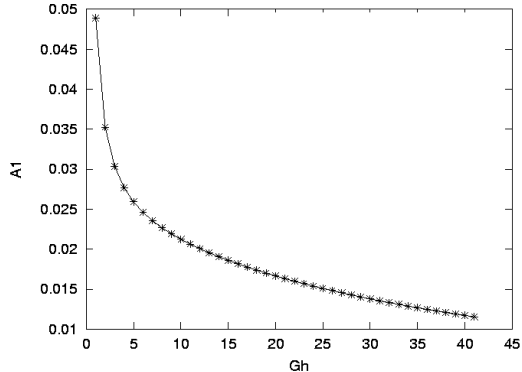}
\caption{Asymmetry ${\cal A}_1$ as a function of the width of the scalar 
$\Gamma_\phi$ in GeV.}
\label{asmwid}
\end{center}
\end{figure}
Fig.~\ref{asmwid} shows the effect on the asymmetry ${\cal A}_1$ of changing 
the width. Thus it is clear that the asymmetries and hence the sensitivities 
will decrease with increasing width. This is borne out by a study of
the blind regions in the various $x_i$--$y_j$ planes analogous to 
our earlier anaylsis, for different values of the widths of the scalar. 

\begin{figure}[htb]
\includegraphics*[scale=.7] {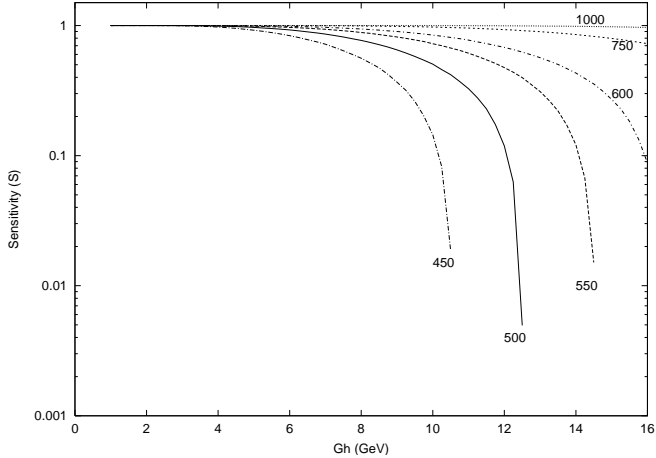} 
\caption{$S_1$ as a function of width of the scalar $\Gamma_\phi$
for different luminosities for the chosen MSSM point and the ideal
backscattered photon spectrum.
}
\label{s1width}
\end{figure}

Next we also study the maximum width of the scalar upto which we can 
discriminate between the SM and the MSSM point at $95\%$ C.L. To that end we 
emply the following procedure. It is clear that the 
SM and the model point chosen will not be confused with
each other if the value of the asymmetry expected for the SM and that
for the model point chosen do not overlap at $95$ C.L. We
generate normally distributed random numbers centered at the
asymmetry corresponding to the SM and take 1 $\sigma$ fluctuation of the
SM asymmetry as the standard deviation.  Let $N_0$ number of generated points.
Let $N_1$ denote the  number of points for which the asymmetry value lies 
within fluctuation expected at $95\%$ C.L. for the expectation of the chosen 
point.  Now probability  ${\cal P}$ of confusing SM with this point  at 
$95\%$ is  given by $N_1/N_0$.  Probability  $P_0$  that $95\% $ C.L. 
intervals of the SM and example point just touch is of course 0.025.  
In this case if  we define  
$$ S_1 = 1- \frac{{\cal P}}{{\cal P}_0},$$
it is easy to see that for $1> S_1 > 0$ the $95\%$ C.L. intervals of the 
SM asymmetry and that expected at the example point do not overlap.
Thus for this case a clear discrimination between the example point and 
the SM possible.  $S < 0$ implies that no such discrimination is possible.
$S_1$ can thus be used quite effectively as a measure of possible 
discrimination. Of course, ${\cal P}$ is dependent on the  angular cut as
well as the chosen
chosen.  We choose the one that gives the smallest $\cal P$
and then plot $S_1$  for different ${\Gamma_\phi}$ and ${\cal L}$.
This is shown in Fig.~\ref{s1width}. The choice of the beam energy and the
cut off angles are the same as used in the earlier analysis.
The figure shows that for a luminosity of 600 fb$^{-1}$ we can
distinguish SM and the chosen MSSM point, with high sensitivity upto 
$\Gamma_\phi = 14$ GeV. Note that this compares well with maximum  width 
expected for a heavy MSSM scalar. 

In view of the rather large effects on the asymmetries and the luminosities
of using the more realistic spectra it is necessary to make a similar study
in that case as well, after optimising the choice of energy and the cutoff
angle $\theta_0$.

\section{Conclusions}
Thus to summarise, we have Studied  $\gamma \gamma \rightarrow \phi \rightarrow
t\bar t$; $\phi$  being a scalar with indefinite CP parity. We looked at the 
process $\gamma \gamma \rightarrow t \bar t \rightarrow l^\pm X$,
where the $l^+/l^-$ comes from decay of $t/\bar t$. We used the most general,
CP nonconserving  ${\cal V}_{\phi \gamma \gamma}, {\cal V}_{t \bar t \phi}$
vertices.  CP violation in these vertices can give rise to net polarisation 
asymmetry for the $t$.  We used the angular distribution for the decay $l$ 
coming from the $t$  as an analyser of $t$ polarisation and hence of CP 
violation in the Higgs sector. We performed our  studies in a model 
independent way by parametrizing the $ {\cal V}_{\phi \gamma \gamma},  
{\cal V}_{t \bar t \phi}$ vertices in terms of form factors.
We showed that decay lepton angular distribution is insensitive
to any anomalous part of the $tbW$ coupling $f^{\pm}$ to first order.
As a result it can be a faithful analyser of the CP violation of the
production process. We constructed  combined asymmetries involving  the
initial lepton (and hence the laser photon) polarisation and the decay lepton
charge. We showed that these can put limits on $CP$ violating combinations,
of the form factors, $y$'s, when only two combinations are varied at a time.
By taking an example MSSM point. We showed that indeed the constructed
asymmetries have sensitivity to CP violation exepected at loop level in
the Higgs sector of the MSSM. We further studied the effect of taking a 
more realistic spectrum for the backscattered laser photon including the
nonlinear effects as well as the effect of an increase in the width of the
scalar. We developed a measure $S_1$ to gauge the ability of the asymmetries 
to discriminate between the SM and our chosen MSSM point, if the scalar were to
have larger width, keeping all the other form factors the same. We were able to
show that with a luminosity of 600 fb$^{-1}$ we can discriminate between the
SM and the chosen MSSM point with high sensitivity upto $\Gamma_\phi = 14$ GeV.

\noindent 
{\bf Acknowledgements} \\
It is a pleasure to  thank T. Matsui, Y.Fujii and R.~Yahata for the impeccable 
organisation of the conference in this beautiful place, which provided a
wonderful backdrop for the very nice/useful  discussions that took place. 
I would like to acknowledge financial support of JSPS which made the 
participation possible. Thanks are also due to the DESY  Theory group for
the hospitality  where  part of this work was carried out. The  
work was partially supported  by the Department of Science and Technology,
India, under project no. SP/S2/K-01/2000-II.

\end{document}